\documentclass[prl,twocolumn]{revtex4}
\usepackage{graphicx,amssymb}
\usepackage{graphicx}
\usepackage{rotating}
\usepackage{dcolumn}
\usepackage{amsfonts}
\usepackage{amssymb}
\usepackage{amsmath}
\usepackage{latexsym}
\usepackage{epsfig}
\usepackage{dsfont,eucal,mathrsfs}

\renewcommand{\mathbf}{\boldsymbol}
\renewcommand{\mathcal}{\mathscr}

\begin{document}

\title{Hub Synchronization in Scale-Free Networks}
\author{Tiago Pereira}
\affiliation{Centro de Matem\'atica, Computa\c{c}\~ao e Cogni\c{c}\~ao \\
Universidade Federal do ABC, Santo Andr\'e, Brasil}

\date{\today}

\begin{abstract}

Heterogeneity in the degree distribution is known to suppress global
synchronization in complex networks of symmetrically coupled oscillators.  
Scale-free networks display a great deal of heterogeneity, containing a few nodes, termed hubs, that are highly connected,  while most nodes receive only a few 
connections. Here, we show that a group of synchronized nodes may appear 
in scale-free networks: hubs undergo a transition to synchronization 
while the other nodes remain unsynchronized.  This general phenomenon can occur 
even in the absence of global synchronization. 
Our results suggest that scale-free networks may have evolved to complement
various levels of synchronization.
\end{abstract}

\maketitle

The last decade has witnessed a tremendous growth of interest in various 
kinds of collective dynamics in networks with complex structures, ranging from 
physical, biological  to social and engineering systems \cite{Bullmore,Kurths,Pecora, Lai,Takashi,Zhou,Ep,Parkinson,Murilo,Ott}. 
Real-world complex systems have been modeled as networks of interacting nodes. 
Synchronized activities have a major impact on the network with important fitness 
consequences to all nodes and network functioning. The network structure 
exerts dramatic influence on its synchronization properties \cite{Pecora,Lai,Takashi}.

Recent studies reveal that disparate real-world networked systems 
share important structural features such as the scale-free property 
\cite{Barrat,Albert}. Scale-free networks are characterized by a 
high level of heterogeneity in the node's degree -- the 
number of connections of a node. Such networks contain a few high-degree nodes, 
termed hubs, while most nodes receive only a 
few connections. The hubs serve specific purposes within their networks, 
such as regulating the information flow and providing resilience during attacks. They severely affect the 
dynamical processes taking place over scale-free networks, 
particularly the emergence of global synchronized motion  
\cite{Pecora,Lai,Takashi}. 

Heterogeneity in the degree distribution may lead to a hierarchical 
transition towards global  synchronization, with hubs synchronizing first, 
followed by the low-degree nodes \cite{Zhou}. In large scale-free 
networks, however, the heterogeneity inhibits global synchronization \cite{Takashi}. 
This turns out to be a desirable property, since in most real-world networks 
where synchronization is relevant, global synchronization 
can be related to pathological activities, such as epileptic seizures \cite{Ep} and 
Parkinson's disease \cite{Parkinson} in 
neural networks. The study of collective behavior apart from 
global synchronization is thus of substantial interest.

In this letter, we show a  general cluster synchronization in scale-free 
networks -- only the hubs undergo a transition to synchronization even in the absence of 
global synchronization. Interestingly, the very heterogeneity that
may prevent global synchronization is the primary ingredient of hub synchronization. 
We provide conditions for the onset of hub synchronization and 
determine the persistence under small perturbations. 
One direct consequence of our theoretical analysis is that hub synchronization is 
both dynamically and structurally stable, thus, allowing the network
to function in a flexible and robust way.

Our approach is to introduce nonlinear dynamics on each node and then 
perform stability analysis to determine when the hubs 
synchronize. From the point the view of stability, reasonable arguments show  
that the network dynamics acts as a small noise-like coupling. Hence, the 
linear stability of the synchronized hubs is maintained. Later on, in the 
large size limit, we provide a  rigorous treatment on the linear stability problem. 
Our analysis is based on the new results of the theory differential equations 
and spectral graph theory. 

We consider a network compose of $n$ nodes, and label the nodes according to 
their degrees $k_1 \le k_2 \le \cdots \le k_n$, where $k_1$ and $k_n$ denote the minimal  and 
maximal node degree,  respectively. Hence, the $i$th node has degree  $k_i$.
A scale-free network is characterized  by the degree distribution $P(k)$, the 
probability that a randomly chosen node within the network has degree $k$, that  follows
a power-law
$
P(k) = c k^{-\gamma}, 
$
for   $k_1 \le k_i \le k_n$,   where c is the normalization factor. The degree
distribution is normalizable for $\gamma > 1$, and for large $k_n$ we have  
$c \approx (\gamma -1 )k_1^{\gamma-1}$. The mean degree $
\langle k \rangle $ attains a finite limit for 
large $k_n$ provided $\gamma>2$. We consider only connected networks with well 
defined mean degree, that is, $ \gamma > 2$.

The dynamics of a general network of $n$ identically coupled elements is described by 
\begin{eqnarray}
\dot{ x }_i &=& {F}({ x}_i) + \frac{\alpha}{k_n} \sum_{j=1}^{n} A_{ij} [{ E}({ x}_j) - { E}({ x}_i)], 
\label{eq1}
\end{eqnarray}
\noindent
here $x_i \in \mathbb{R}^m$ is the $m$-dimensional vector describing the state
of the $i$th node (node with degree $k_i$), $F: \mathbb{R}^m \rightarrow \mathbb{R}^m$ 
governs the dynamics of the individual oscillator and is assumed to be smooth,  
${E} : \mathbb{R}^m \rightarrow \mathbb{R}^m$ is the coupling function (without loss of 
generality assumed to be a constant  matrix), $\alpha$ is the normalized overall coupling strength \cite{couplingStrength}, 
and $A$ is the adjacency matrix. $A$ encodes 
the topological information of the network, defined as $A_{ij}=1$ if nodes $i$ and $j$ 
are connected and $A_{ij}=0$ otherwise. Note that $A$ is symmetric, and by definition $k_i = \sum_j A_{ij}$.  

We wish to show that a group of oscillators having nearly the same number of 
connections as the main hub may display a synchronized motion. 
Consider $\xi_i = x_n - x_i$, thus, synchronization is possible between the nodes 
$i$ and $n$ if $\xi_i \rightarrow 0$. Stability of this synchronized state is determined 
by analyzing the variational equations governing the perturbations, which read
\begin{eqnarray}
\dot{{\xi}}_{i} =  K_i(t; \alpha) {\xi}_i  + \alpha \eta_i,
\label{xi}
\end{eqnarray}
\noindent
where the matrix $K_i(t; \alpha) = [{D} {F}({x_n(t)})  -  \alpha \mu_i E] $ depends continuously on $t$, ${D} {F}$  stands for the Jacobian matrix of $F$, $\mu_i = k_{i}/{k_{n}}$ is the normalized degree, and 
$$
\eta_i  =  \frac{1}{k_n} \sum_j ( A_{ i j} - A_{ n j}) E ({\xi}_j)
$$
is the coupling term. 

Neglecting the coupling term $\eta_i$ the equations governing the evolution 
of the perturbations $\xi_i$ and are decoupled from the other perturbations 
and read
\begin{equation}
\dot{{\xi}}_i = [ {D}{F}(x_{n}(t)) + \alpha \mu_i {E} ] {\xi}_i.
\label{nozeta}
\end{equation}

We now assume that Eq. (\ref{nozeta}) is Lyapunov regular and that its 
fundamental matrix is integrally separated \cite{commentAssump}.  
The stability of the zero solution of Eq. (\ref{nozeta}) is determined by 
its largest Lyapunov exponent $\Lambda(\alpha \mu_i)$, which can be regarded as 
the {\it master stability function} of the system \cite{Pecora,Lai}. The 
perturbation $\xi_i$ is damped out  if  $\Lambda(\alpha \mu_i) < 0$.

For many  widely studied oscillatory systems the master stability function 
$\Lambda (\alpha \mu_i)$ is negative in an interval $\alpha_1 < \alpha \mu_i <  
\alpha_2$ for general coupling function $E$ \cite{Pecora,Lai}. The perturbation 
$\xi_i$ is damped out if $\alpha_1 <  \alpha \mu_i < \alpha_2$. Moreover, 
normalization imposes $\mu_n =1$ and $\mu_1 \propto k_n^{-1}$, hence, as 
$k_n$ increases, $\mu_1$ converges to zero. Not only $\mu_1$, but most of 
the normalized degrees $\mu_i$ will converge to zero. Therefore, it 
will be impossible, for large $k_n$, to have $\alpha_1 <  \alpha \mu_i < \alpha_2$ 
for all $i=1,2,  \cdots, n$. Hence, in the thermodynamic limit no stable global 
synchronization is possible in scale-free networks. 

Now take $\alpha$ in the stability region. Then, the  state $x_n = x_{n-1}$ is 
linearly stable. This is true as long as we can neglect the coupling term $\eta_i$.  
Under the effect of $\eta_i$ local mean field arguments show that $x_n \approx 
x_{n-1}$ is stable. The argument goes as follows. If $\Lambda(\alpha \mu_{n-1} ) 
< 0$, we guarantee the linear stability of  $\xi_{n-1}$. Moreover, if the remaining 
oscillators are not synchronized, the coupling term  $\eta_{n-1}$ can be  viewed 
as a small coupling noise, as long as the signals $x_i$  are uncorrelated, with 
$\alpha$ fixed and $k_n$ large \cite{localmeanfield}. Results from ordinary 
differential equations state that the linear stability is maintained under small 
perturbations \cite{LyapSt,EDO1}.  Therefore, if at $t=0$ we have  $x_n(0) - 
x_{n-1}(0) \approx 0$, then for all  $t \ge 0$ it yields  $x_n(t) - x_{n-1}(t) \approx 0$.

These arguments cannot be applied to low-degree nodes. 
The reason is that to set the low-degree nodes into the stability region we must have  
$\alpha \mu_1 \approx \alpha_1$, requiring  $\alpha$ to be as large as $k_n$. Hence, 
the coupling term $\alpha \eta_i$ cannot be made small for low degree nodes.

The mean field arguments also hold for  {\it correlated scale-free networks}. 
The node correlation does not play a major role to the onset of hub synchronization. 
For instance, the  Barab\'asi-Albert (BA) scale-free model is known to present 
finite size node correlation, hubs are likely connected \cite{Albert}.  If we rewire the 
connections between the hubs, connecting the hubs with the low degree 
nodes, the mean field argument is still valid, that is,  hub synchronization still 
takes place. 

We illustrate this phenomenon with numerical 
experiments. We generate a Barab\'asi-Albert (BA) scale-free 
network with $3\times 10^3$ nodes and $m=3$ \cite{Albert}.  The network has 
largest degrees $k_n=k_{n-1}=165$. Each node $x_i$ is 
modeled as a R\"ossler oscillator, for $x_i = (x_{1i}, x_{2i}, x_{3i})^T$ we 
have $F(x_i) = (x_{2i} - x_{3i}, x_{1i} + 0.2 x_{2i}, 0.2 + x_{3i}(x_{1i} - 7  ) )^T$. 
We consider $E$ to be a projector in the first component, i.e.,  
$E(x,y,z)^T = (x,0,0)^T$. The master stability function  $\Lambda (\alpha) $ has 
a stability region for $ \alpha \in (\alpha_1, \alpha_2)$ with $\alpha_1 \approx 0.13$ 
and $\alpha_2 \approx 4.55$. Global synchronization in this network is 
impossible \cite{commentSync}.

\begin{figure}[!htb]
\includegraphics[scale=0.6]{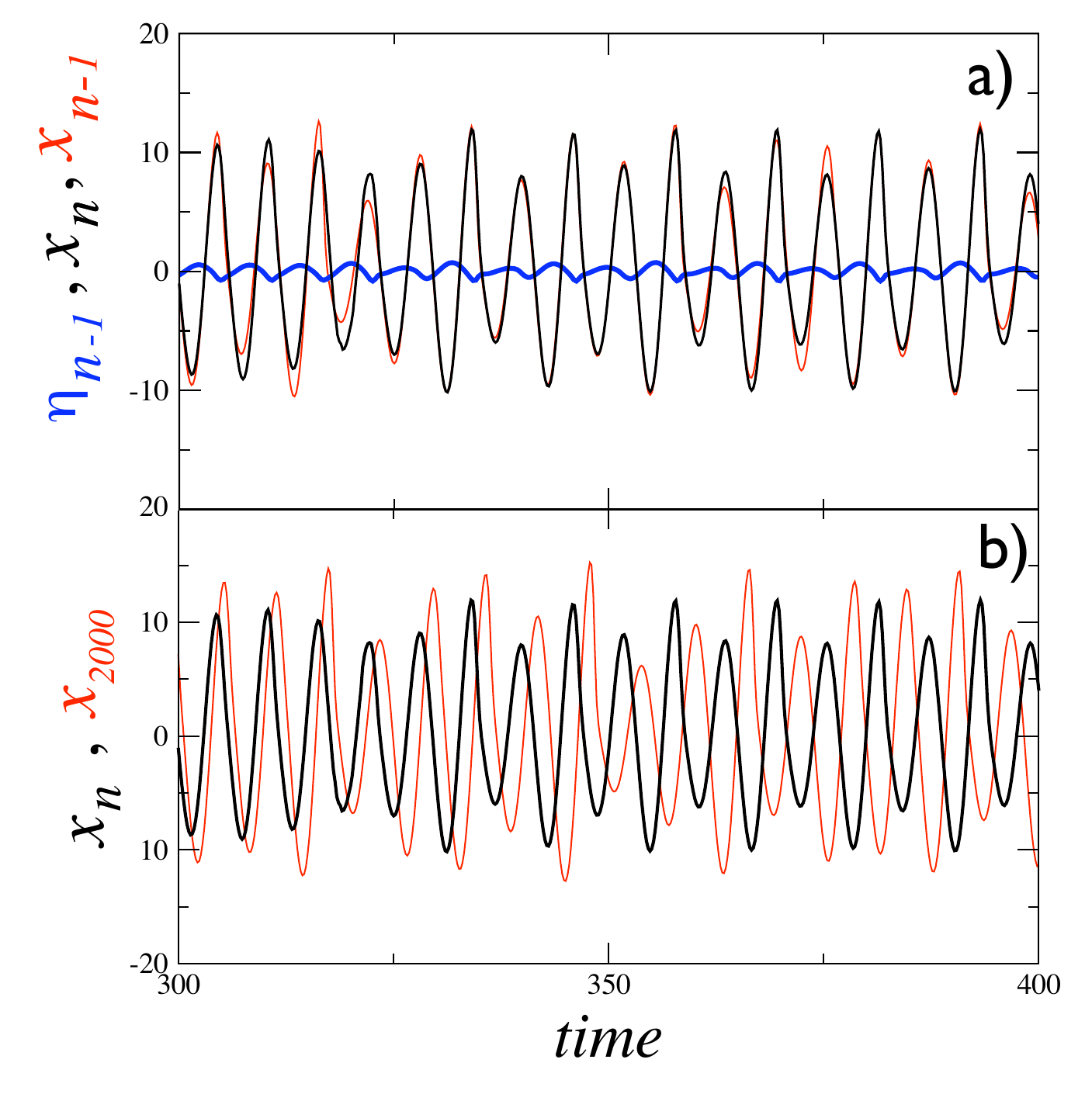}
\caption{[Color online] Hub synchronization in a BA scale-free network 
of 3000 coupled R\"ossler oscillators with coupling parameter $\alpha=0.3$. 
a) Time series of the largest hub $x_n$ (full line) and the second largest  
$x_{n-1}$ (light gray line). The coupling term $\eta_{n-1}$ (bold line) spoiling the stability 
of the hub synchronization is small as predicted by the local mean field arguments.
b) Time series of the largest hub $x_n$ (full line) and of a low-degree node $x_{2000}$ (light gray line)
The corresponding  node degrees are $k_n = k_{n-1}=165$ and $k_{2000}=3$.}
\label{Fig1}
\end{figure}

For $\alpha = 0.30$ we have observed the hub synchronization 
$x_n \approx x_{n-1}$. In Fig. \ref{Fig1}(a) the time series
$x_n$ is depicted in full line while $x_{n-1}$ is depicted in light gray line and
$\eta_{n-1}$ in bold line. Fig. \ref{Fig1} (a)
shows that the local mean field approximation on $\eta_{n-1}$ indeed
holds, as shown in the times series  $x_{n-1}  \approx x_{n}$.
In Fig. \ref{Fig1}(b), the differences $\xi_{n-1} = x_{n-1}-x_n$ is depicted in bold line
while  $\xi_{2000} = x_{2000} - x_n$ in full line. Clearly $\xi_{n-1} \approx 0$
whereas $\xi_{2000}$ presents large fluctuations.

\noindent

All this reasoning can be set into a rigorous frame in the thermodynamic limit, 
for uncorrelated scale-free networks.  To tackle the problem let us introduce 
$\zeta_i(t) = x_i(t) -  s(t) $, where $s(t)$ is a given typical trajectory of 
$\dot{x} = F(x)$. Consider  $\mathbf{\zeta} = ({\zeta}_1, {\zeta}_2, \ldots , {\zeta}_n)^T$ 
and $\mathbf{\mu} = diag (\mu_1 , \mu_2 , \ldots, \mu_n)$. Hence, 
$\mathbf{\zeta} \in \mathbb{R}^{mn}$. The variational equations of the perturbations 
$\mathbf{\zeta}$ can be written in a convenient block form
\begin{equation}
\dot{\mathbf{\zeta}} =  \mathbf{\Omega}(t ; \alpha) \mathbf{\zeta} + \alpha \mathbf{B} \mathbf{\zeta}
\label{eq3}
\end{equation}
where $\mathbf{\Omega}(t; \alpha) = I_n \otimes D F(s(t)) - \alpha \mathbf{\mu} \otimes E$, 
with $\otimes$ standing for the Kronecker product, and  
$\mathbf{B}  = k_{n}^{-1} A \otimes E$ is the coupling among the variational equations. 
We shall demonstrate that for large scale-free network with $\gamma>2$, the term coupling term can be made arbitrarily small. 

According to the aforementioned arguments $ \mathbf{\Omega}(t ; \alpha)$ splits into 
independent blocks as in Eq. (\ref{nozeta}). By choosing a fixed $\alpha$ such that nodes 
with degree larger than $k_{n-\ell}$ have their perturbations damped out, we 
guarantee that $\ell$ nodes display a synchronous behavior with the main hub $x_n$. 
In other words, $\mathbb{R}^{nm} = U \oplus S$, where
$U$ and $S$ respectively the unstable and stable spaces, clearly dim$(U) = (n- \ell ) m$
and dim$(S)  = \ell m$. Notice that on the subspace $S$ all Lyapunov exponents are 
negative. 

It remains to show that the coupling term can be made as small 
as one wishes whenever $k_n$ is large enough. Thus, results of qualitative theory 
of ordinary differential equations guarantee that the linear stability is not affected by 
small continuous perturbations \cite{Stability}.

By our hypothesis on the symmetry 
of the matrix $A$ the spectral theorem  guarantees that 
$$
A = N J N^{-1}.
$$
where $N$ is an orthogonal matrix and  $J=diag(\lambda_1, \lambda_2, \ldots, \lambda_n)$ 
is the matrix of the eigenvalues of $A$ ordered according to their magnitudes 
$\lambda_1 \le \lambda_2 \le \cdots < \lambda_n$.  

We endow the vector space $\mathbb{R}^{mn}$ with the norm $\| \cdot \| _{*}$ such that for 
$u \in \mathbb{R}^{mn}$  we have $\| u \| _{*} = \| N \otimes I_m  u \| _{\infty}$,
where $\| u \|_{\infty} = \sup_i |u_i |$ for $i = 1,2, \cdots, nm$. We also make 
use of the  induced matrix norms.
Now we claim that  given $\delta > 0$ there exists $K$ such that for all 
$k_n > K$ we have
$$
 \|  \mathbf{B} \|_*  < \delta.
$$
Indeed, by using the  induced matrix norm we can obtain bounds in terms of the 
largest eigenvalue of $A$. We postpone the technical details and go directly to 
the result which reads 
$\|  A \otimes E \|_{*} \le \lambda_n \| E \|_{\infty}$.

Under mild conditions \cite{commentEigen} the largest eigenvalue  of a scale-free 
network scales almost surely as $\lambda_n = k_n^{\beta}$, where depends 
on $\gamma$. We have two distinct cases: 
$(i)$ $\beta = 3 - \gamma $ for $2 < \gamma < 2.5$; and
$(ii)$ $\beta = 1/2 $ for $\gamma > 2.5$. Putting all estimates together yields
\begin{equation}
\|  \mathbf{B} \|_*  \propto \frac{1}{k_n^{ 1 - \beta}}.
\end{equation}
Hence, for $k_n$ large enough our claim follows. 
 
This analysis is grounded on the fact that $\lambda_n/k_n \rightarrow 0$. This is also 
the case for {\it correlated scale-free networks} \cite{Ott}, whenever the correlations 
preserve the scale-free character. These moderate correlations are immaterial for 
hub synchronization,  as  finite size correlation in the BA scale-free model. 
  
In summary,  we analyzed a general phenomenon in the synchronization of 
large scale-free networks, namely, the synchronization of hubs even when 
the entire network is out of synchrony. Our theoretical analysis provides insights into 
further generalizations for the master stability function. The stability analysis of the 
synchronous hubs  can be tailored to the master stability function and the coupling 
term due to the underlying network dynamics. We have shown that for large scale-free 
networks the coupling term can be controlled, effectively acting  as a small noise-like 
perturbation on the hubs.

Hub synchronization has counterintuitive effects. For example, the hubs do 
not need to be directly connected to synchronize. Remarkably, when the hubs 
synchronize, the  low-degree nodes are out of synchrony; these nodes, however,   
are responsible for mediating the exchange of information between the hubs. 
This seems to challenge our understanding of the role of synchronization 
in the exchange of information within complex networks  \cite{Murilo}.

We believe that our findings provide strong evidence that incomplete, hub-driven, 
synchronization may be at least as important and persistent in real-world networks as  other forms of synchronization and collective behaviors previously examined in the literature.

The author is in debt with Rafael D. Vilela,  Alexei M. Veneziani, Murilo S. Baptista 
and Adilson E. Motter for illuminating discussions, a detailed and critical reading of the manuscript.
This work was partially supported by CNPq grant 474647/2009-9.


\begin{thebibliography}{99}



\bibitem{Bullmore} E. Bullmore and O. Sporns,  Nature Neurosc. {\bf 10}, 186 (2009);
V.M. Eguiluz, D.R. Chialvo, G.A. Cecchi, M. Baliki, A.V. Apkarian, Phys. Rev. Lett. {\bf 94}, 018102 (2005).
 
\bibitem{Kurths} A. Arenas, A. D\'iaz-Guilera, J. Kurths, Y. Moreno and C. Zhou,
Phys. Rep.  {\bf 469}, 93-153 (2008).

\bibitem{Pecora} L.M. Pecora  and T.L. Carroll, Phys. Rev. Lett. {\bf 80}, 2109 (1998); 
 M. Barahona and L.M. Pecora, Phys. Rev. Lett. {\bf 89}, 054101 (2002).


\bibitem{Lai} L. Huang, Q. Chen, Y.C. Lai, and L.M. Pecora, Phys. Rev. E {\bf 80}, 036204 (2009).


\bibitem{Takashi} T. Nishikawa, A.E. Motter, Y.C. Lai, and F.C. Hoppensteadt,
Phys. Rev. Lett. {\bf 91} (2003) 014101; A.E. Motter, C. Zhou, and J. Kurths, Phys. Rev. E {\bf 71}, 
016116 (2005).


\bibitem{Zhou} C. Zhou and J. Kurths, Chaos {\bf 16}, 015104 (2006); 
J. Gomez-Gardenes, Y. Moreno, and A. Arenas, Phys. Rev. Lett. Ê{\bf 98}, 034101 (2007); 
D.-S. Lee,  Phys. Rev. E 72, 026208 (2005).


\bibitem{Ep} F. Mormann, T. Kreuz, R. G. Andrzejak, P.
  David, K.  Lehnertz, and C. E. Elger, Epilepsy Research {\bf 53}, 173 (2003).

\bibitem{Parkinson} P. Tass, M.G. Rosenblum, J.Weule, et al., Phys. Rev. Lett. 81, 3291 (1998). 


\bibitem{Murilo}  Baptista M. S.,  de Carvalho J. X., Hussein M. S., PLoS ONE  {\bf 3}, e3479 (2008). 


\bibitem{Ott} J. G. Restrepo, E. Ott, and B. R. Hunt, Phys. Rev. E {\bf 76}, 056119 (2007).




\bibitem{Barrat} A. Barrat, M. Barthelemi, A. Vespegnani, {\it Dynamical Processes 
on Complex Networks}, Cambridge University Press (2008); M. Newman, A.-L. Barab\'asi, 
and D. J. Watts, {\it The structure and dynamics of networks}, Princeton University Press (2006). 


\bibitem{Albert} Albert R., Jeong H. and Barab\'asi A.-L. , Nature {\bf 406} (2000) 378;
Albert  R.  and Barab‡si A.-L., Rev. Mod. Phys. {\bf 74}, 47 (2002).


\bibitem{couplingStrength} The choice of normalized coupling is immaterial. 
The choice does not play any role in the analysis.
Note that we use the same normalization 
for all nodes, so we could have written $\sigma = \alpha / k_n$, 
as is usually done in the literature. 


\bibitem{commentAssump} In our context these are natural assumptions. Lyapunov regularity 
basically assures that the Lyapunov exponents exist. The integral separation 
is a generic property in the space of continuous bounded matrix valued functions. See 
\cite{LyapSt} for a detailed discussion.


\bibitem{LyapSt} D.J. Estep and S. Tavener (Eds.) {\it Collected Lectures on the Preservation of Stability 
under Discretization}, SIAM (2002).


\bibitem{localmeanfield} 
Without attempt at rigor, the local mean field argument is the following. 
First remember that 
$\eta_{n-1}  = k_n^{-1} \sum_j ( A_{ (n -1) j} - A_{ n j}) E ({\xi}_j)$, 
since the oscillators are chaotic and unsynchronized (at least for small values of $\alpha$)
once can think of $\xi_j$ as identically distributed random numbers.
For $k_n \gg 1$,  by the center limit theorem $\eta_i = O(k_n^{-1/2})$.


\bibitem{EDO1} L.  Barreira  and C. Valls, {\it Stability of Nonautonomous 
Differential Equations}, Springer-Verlag Berlin Heidelberg (2008).  




\bibitem{commentSync} The stability of global synchronization is formulated in 
terms of the spectrum of laplacian matrix. Let $L$ be the laplacian matrix of the graph. 
The spectrum of $L$ is real 
and can be ordered as $0=\gamma_1 \le \gamma_2 \le \cdots \le \gamma_n$.
Global synchronization is possible if $\gamma_n/ \gamma_2 < \alpha_2 / \alpha_1$, see \cite{Pecora,Takashi}
for details. For this network we have, $\gamma_n/\gamma_2 \approx 180$, while 
$\alpha_2 / \alpha_1 \approx 35$.


\bibitem{Stability}  The unique solution of the homogeneous 
part of Eq. (\ref{eq3}) can be written in terms of the principal matrix 
$
\mathbf{\zeta}(t) = T (t, s) \mathbf{\zeta}(s). 
$ 
Moreover, under our hypotheses the operator $T(t,s)$
admits a dichotomy being exponentially stable on the 
subspace $S$ \cite{EDO1}. Since $\|Ê \mathbf{B} \|_* <  \delta $, 
it follows that the Lyapunov exponents of the perturbed equation remain
negative \cite{LyapSt}. 

\bibitem{commentEigen} The conditions require  $k_n$ to grow faster than 
some powers of $\log n$.  See \cite{Comb} for further details. These conditions are all 
natural for large scale-free graphs,  almost surely, it holds $k_n \propto n^{1/(\gamma-1)}$. 

\bibitem{Comb} F. R. K. Chung and L. Lu, {\it Complex Graphs and Networks}, American Mathematical Society (2006).



\end{thebibliography}
\end{document}